\documentclass[12pt, preprint]{aastex}

\shorttitle{Sirius B in the Mid-Infrared}
\shortauthors{Skemer et al.}

\begin{document}

\title{Sirius B Imaged in the Mid-Infrared: No Evidence for a Remnant Planetary System\footnote{Based on observations obtained at the Gemini Observatory, which is operated by the Association of Universities for Research in Astronomy (AURA) under a cooperative agreement with the NSF on behalf of the Gemini partnership: the National Science Foundation (United States), the Science and Technology Facilities Council (United Kingdom), the National Research Council (Canada), CONICYT (Chile), the Australian Research Council (Australia), CNPq (Brazil) and CONICET (Argentina)}}

\author{Andrew J. Skemer, Laird M. Close}
\affil{Steward Observatory, Department of Astronomy, University of Arizona, Tucson, AZ 85721}

\begin{abstract}
Evidence is building that remnants of solar systems might orbit a large percentage of white dwarfs, as the polluted atmospheres of DAZ and DBZ white dwarfs indicate the very recent accretion of metal-rich material. \citep{2010ApJ...722..725Z}.  Some of these polluted white dwarfs are found to have large mid-infrared excesses from close-in debris disks that are thought to be reservoirs for the metal accretion. These systems are coined DAZd white dwarfs \citep{2007ApJ...662..544V}

Here we investigate the claims of \citet{2008A&A...489..651B} that Sirius B, the nearest white dwarf to the Sun, might have an infrared excess from a dusty debris disk.  Sirius B's companion, Sirius A is commonly observed as a mid-infrared photometric standard in the Southern hemisphere.  We combine several years of Gemini/T-ReCS photometric standard observations to produce deep mid-infrared imaging in five $\sim$10$\micron$ filters (broad N + 4 narrowband), which reveal the presence of Sirius B.  Our photometry is consistent with the expected photospheric emission such that we constrain any mid-infrared excess to $\lesssim$10\% of the photosphere.  Thus we conclude that Sirius B does not have a large dusty disk, as seen in DAZd white dwarfs.
\end{abstract}

\section{Introduction}
Since its discovery in 1844 \citep{1844MNRAS...6R.136B}, Sirius B has been a tantalizing object.  While its close proximity to the Sun makes it ideal for detailed study, its binary companion, Sirius A, is the brightest star in the night-sky, complicating observations of Sirius B.  With a $\Delta$ mag of $\sim$10 between Sirius A and B for optical and longer wavelengths, the most successful observations of Sirius B have been space-based.  \textit{Hubble Space Telescope (HST/STIS)} spectra of Sirius B have led to the accurate determination of its effective temperature ($T_{\rm eff}=$25,193 K) and mass \citep[0.978 $M_{\sun}$;][]{2005MNRAS.362.1134B}, \textit{Hipparcos} parallax measurements of Sirius A place the system 2.64 pc from the Sun \citep{1997A&A...323L..49P}, and astrometric monitoring has determined Sirius B's orbital period (50.090 yrs), semi-major axis (7.500" or 19.8 AU), and eccentricity \citep[0.5923;][]{2001AJ....122.3472H}\footnote{http://ad.usno.navy.mil/wds/orb6.html}.

Increasingly, ground-based adaptive optics (AO) systems are becoming adept at high-contrast imaging, usually with the goal of discovering faint planets/brown dwarfs near their bright host stars.  Imaging a faint white dwarf like Sirius B around a bright, main-sequence A star like Sirius A is a similar problem.  Recently, \citet{2008A&A...489..651B} used the ESO 3.6 meter telescope, along with the ADONIS AO system, to image Sirius B in the JHKs filters, which, before this work, were the longest wavelength photometric measurements of Sirius B.  The \citet{2008A&A...489..651B} measurements showed a small (1.7$\sigma$) Ks-band excess when compared to the models of \citet{2006AJ....132.1221H}.  Similar near-IR excesses are present in the spectra of dusty white dwarfs, which are characterized by large excesses in the mid-infrared \citep{2007ApJ...662..544V}.

The first infrared excess around a white dwarf (G29-38) was found by \citet{1987Natur.330..138Z}, and was initially thought to be a brown dwarf.    This hypothesis was ruled out by optical/near-IR pulse monitoring that suggested a disk-geometry source of the excess \citep{1990ApJ...357..216G,1991ApJ...374..330P}.  Subsequent spectroscopy using \textit{Spitzer}/IRS revealed the presence of small dust grains \citep{2005ApJ...635L.161R}.  A \textit{Spitzer}/IRS survey found 4 dusty white dwarfs out of their sample of 124 \citep{2007ApJS..171..206M,2007ApJ...662..544V}, and that each of the dusty white dwarfs were of type DAZ (metal-polluted, hydrogen atmosphere), leading \citet{2007ApJ...662..544V} to coin DAZ stars with mid-infrared excesses, DAZd.  The metals in DAZ white dwarfs are expected to settle below the white dwarf photosphere much faster than evolutionary timescales \citep{2006A&A...453.1051K}, thus the presence of metals implies a recent accretion event.  $\sim$1/4 of DA (hydrogen atmosphere) white dwarfs are DAZ \citep{1998ApJ...505L.143Z}, and $\sim$1/3 of DB (helium atmosphere) white dwarfs are DBZ \citep{2010ApJ...722..725Z}, demonstrating that accretion must be a common phenomenon for white dwarfs.  The cause of this accretion, in many if not all cases, is thought to be tidally disrupted asteroid-sized objects from a remanent debris disk/planetary system \citep{2008AJ....135.1785J,2010ApJ...722..725Z}.  

We note that Sirius B is a DA white dwarf, but determining if it is metal polluted (DAZ) is complicated by the difficulty of taking high-resolution spectra of Sirius B from the ground.  This problem is exacerbated by the fact that Sirius B's high temperature, \citep[25,193 K][]{2005MNRAS.362.1134B} impedes the detection and interpretation of photospheric metals \citep{2006A&A...453.1051K,1995ApJ...454..429C}.  Sirius B's high temperature also has implications for its potential to host a debris disk.  While typical DAZd white dwarfs have temperatures ranging from T$\approx$7,000-15,000 K \citep{2009ApJ...694..805F}, Sirius B's high temperature would sublimate dust out to its tidal truncation radius \citep{2003ApJ...584L..91J}, precluding tidal disruption of asteroids as the source of dust in the system.  Instead of dust disks, hot white dwarfs can have metal vapor disks.  The first system observed to display these features, SDSS 1228+1040 \citep{2006Sci...314.1908G}, was subsequently found to have a dust disk at larger radii \citep{2009ApJ...696.1402B}. Analogously, a debris disk around Sirius B would be at larger radii than is typical for DAZd white dwarfs, and the dust would be the result of collisions, rather than tidal disruption of asteroids.  Since not all DAZ and DBZ white dwarfs are found to have debris disks, the source of their photospheric pollution might come from larger radii than are easily probed by current mid-infrared instruments (24$\micron$ probes $\sim$120 K dust, which is at $\lesssim$1 AU, assuming radiative equilibrium).  As a result, little is known about the outer regions of white dwarf debris disks.

Spatially resolving a debris disk around a white dwarf would obviously be an important step in understanding the DAZd phenomenon.  As the nearest white dwarf to the Sun, Sirius B would be a prime target for such a search, especially given the claims of \citet{2008A&A...489..651B} that Sirius B might have a substantial mid-infrared excess.  The importance of such a discovery would be magnified by the fact that dust particles $\gtrsim$1 AU from Sirius B would be primarily heated by Sirius A (18 AU in 2005.0), which would potentially make it possible to probe the outer parts of the debris disk at much greater separations than is normally possible.  The maximum radius of Sirius B's disk, based on tidal truncation from Sirius A, would be $\sim1/4$ of the orbital semi-major axis \citep{1994ApJ...421..651A}, which is 5 AU or 2".

In this work, we use Gemini/T-ReCS \citep{1998SPIE.3354..534T} archival observations of Sirius to determine if Sirius B has a strong mid-infrared excess.  As the primary Cohen standard for the Southern hemisphere \citep{1992AJ....104.1650C}, Sirius A is commonly observed as a photometric calibration for a variety of T-ReCS programs.  By co-adding these data, we obtained deep mid-infared images, allowing us to detect the very faint source, Sirius B, at high signal-to-noise, and with the redundancy of independent detections in 5 filters.

\section{Observations and Reductions}
We downloaded all public imaging data of Sirius taken by the Thermal-Region Camera Spectrograph \citep[T-ReCS;][]{1998SPIE.3354..534T} on the Gemini-South Telescope.  We reduced our chop/nod data and removed bad frames using the custom T-ReCS IDL software MEFTOOLS v. 5.0\footnote{http://www.jim-debuizer.net/research/}.

Most of the data were taken with T-ReCS oriented at a position angle (PA) of 0$^{\circ}$ (North is up and East is left on the detector).  We removed the few images where PA was not 0$^{\circ}$, and we also removed all data taken in 2007 or later, because the expected orbital position of Sirius B was near detector artifacts.

Without prejudice to observing conditions or image quality, we registered and coadded the remaining images in each filter, weighting each image by on-source observing time.  All registration was done on Sirius A, as Sirius B is not visible in the individual images.  Hereafter, we refer to these images as the A-aligned images. The total on-source observing time in each filter is listed in Table \ref{Observations}.

\begin{figure}
\center{
 \includegraphics[angle=0,width=0.6\columnwidth]{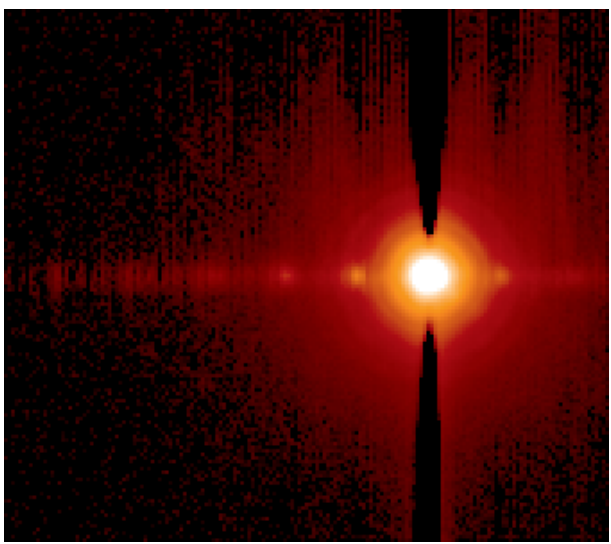}}
\caption{Sirius is commonly observed by Gemini/T-ReCS observers for calibration.  This image combines all Si-2 (8.74$\micron$) data of Sirius taken between  2003, Dec. 30 UT and 2006 Jun. 31 UT, aligned and coadded on Sirius A.  Sirius A's PSF is symmetric, diffraction-limited and unsaturated despite a variety of observing conditions and strategies.  The image is 13" by 18".  North is up and East is left.  Sirius B is not visible in this stretch.
\label{Sirius A image}}
\end{figure}

\begin{figure}
\center{
 \includegraphics[angle=0,width=\columnwidth]{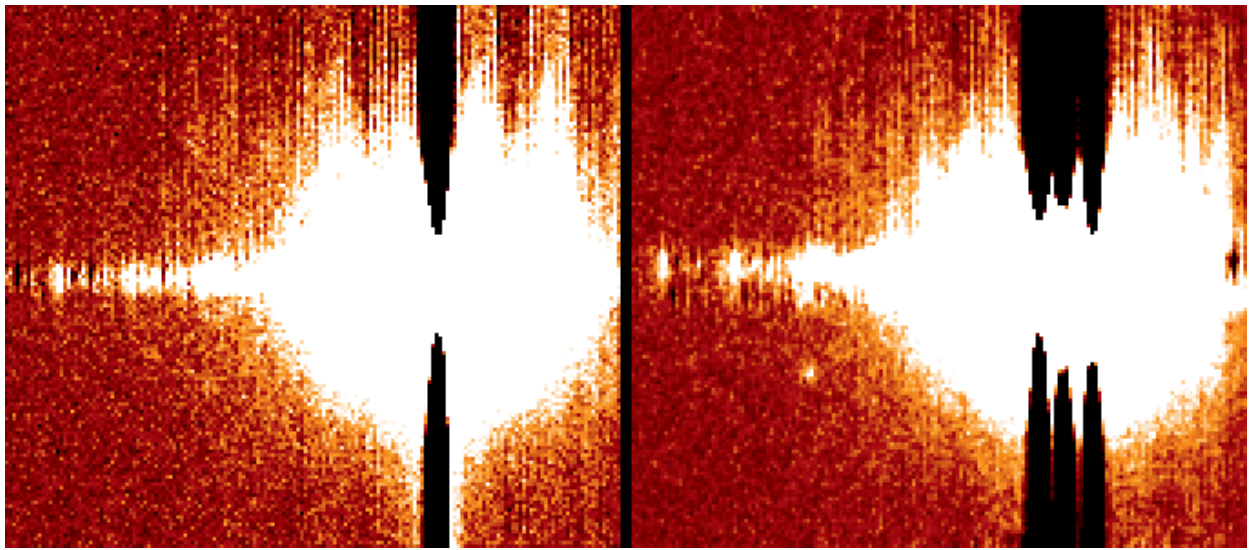}}
\caption{Each image is 13" x 18".  North is up and East is left.  LEFT: Sirius, aligned and coadded on Sirius A in the Si-2 (8.74$\micron$) filter (same as Figure \ref{Sirius A image} but at a different stretch).  A very faint streak is visible in the lower-lefthand quadrant, which is Sirius B smeared out by its orbital motion between epochs ($\sim$3 years).  RIGHT:  The same data, but where each frame is shifted by the calculated change in the position of Sirius B, so that the image is aligned on Sirius B.  Sirius B is now clearly visible at a position angle of 113$^\circ$ and separation of 6.75" (its 2005.0 position).
\label{Sirius B image}}
\end{figure}

Figure \ref{Sirius A image} shows the A-aligned image of Sirius in the Si-2 (8.74$\micron$) filter.  The point-spread function (PSF) of Sirius A is symmetric, diffraction-limited, and unsaturated.  Figure \ref{Sirius B image} again shows the A-aligned image of Sirius (on the left), but this time with a deeper stretch that brings out the background at $\gtrsim$4".  The image shows a very faint streak in the lower-lefthand quadrant, at the approximate orbital positions of Sirius B from 2003-2006.  Because our image has been constructed from data spanning $\sim$3 years, Sirius B's fast orbital motion ($\sim$50 year period) has created an arc across our A-aligned images.

To make B-aligned images, we have to ``de-orbit" the individual images that have been registered on Sirius A.  We use orbital elements from the US Naval Observatory's Sixth Catalog of Orbits of Visual Binary Stars\footnote{http://ad.usno.navy.mil/wds/orb6.html} \citep{2001AJ....122.3472H} and shift each individual image by the calculated orbital ephemeris to stack each image on Sirius B.  The final stacked image is shown on the right of Figure \ref{Sirius B image}.  Sirius B is now clearly visible.

\section{Mid-Infrared Photometry of Sirius B}
We perform PSF-fitting photometry with IRAF/\textit{daophot} \citep{1987PASP...99..191S} using Sirius A (from the A-aligned image) as our PSF.  The $\Delta$ mag between Sirius A and B and fluxes are listed in Table \ref{Observations} for each filter.  The associated errors are the difference between using two sky background annuli (6-9 pixels vs. 9-12 pixels) combined in quadrature with the measured \textit{allstar} errors.  

In the Si-2 and Si-5 filters, the shape of Sirius B's PSF is essentially identical to that of Sirius A, with no signs of extended emission after PSF subtraction.  The Si-3 and Si-4 filters have much lower S/N detections, so that a comparison is difficult.  In the N-band filter, the PSFs look somewhat different, likely because of contamination by detector artifacts, which are stronger in the N-band filter than in the Si filters.  As a result, we perform aperture photometry on our N-band data in lieu of PSF-fitting photometry.  We use 3 different apertures (3, 4 and 5 pixels) to estimate our error, which is large (0.146 mag), due to the detector artifacts.

\begin{figure}
\center{
 \includegraphics[angle=90,width=\columnwidth]{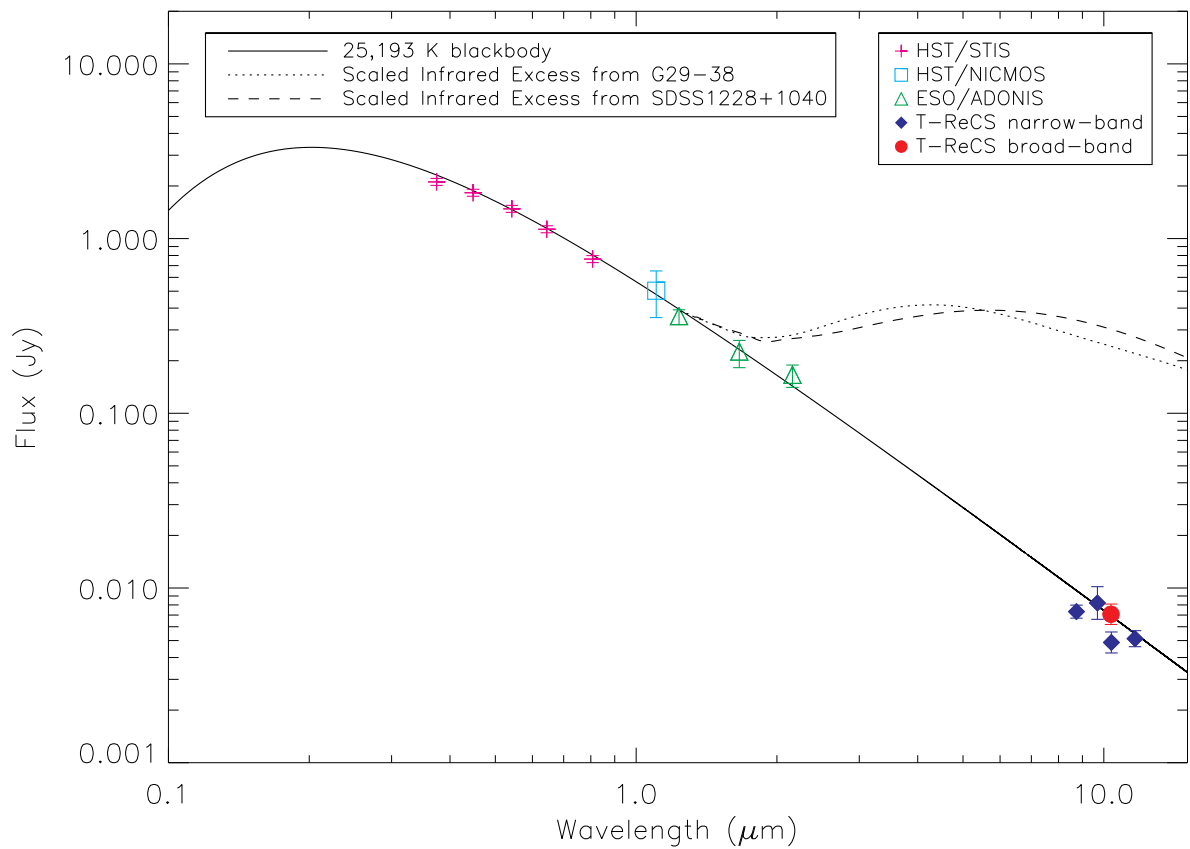}}
\caption{SED of Sirius B with data from HST/STIS (pink plus-signs), HST/NICMOS (teal square), ESO/ADONIS (green triangles), Gemini/T-ReCS narrowband Si filters (filled, blue diamonds) and Gemini/T-ReCS broadband N (filled, red circle).  The solid curve is a blackbody with temperature=25,193K scaled to the STIS V-band flux, the dotted curve is the scaled infrared excess of the prototype DAZd white dwarf, G29-38, and the dashed curve is the scaled infrared excess of the hot white dwarf SDSS 1228+1040.  Our Gemini/T-ReCS data show no evidence of a mid-infrared excess as hypothesized by \citet{2008A&A...489..651B}.
\label{Sirius B SED}}
\end{figure}

Because Sirius A is the primary standard of \citet{1992AJ....104.1650C} for the Southern hemisphere, it has a well-calibrated flux within the Gemini filters\footnote{http://www.gemini.edu/sciops/instruments/mir/MIRStdFluxes.txt} that we can use to flux calibrate Sirius B.  We plot these fluxes on a spectral energy distribution (SED) in Figure \ref{Sirius B SED}, along with HST/STIS photometry \citep{2005MNRAS.362.1134B}, HST/NICMOS photometry \citep{2000PASP..112..827K}, and ESO/ADONIS photometry\footnote{In Figure \ref{Sirius B SED}, the ESO/ADONIS photometry do not appear to show an excess, as claimed by \citet{2008A&A...489..651B}.  Although our blackbody normalization might be slightly different than \citet{2008A&A...489..651B}'s, we note that an excess is similarly not visible in their Figure 2.  Rather, they claim an excess via comparison of their absolute photometry to the models of \citet{2006AJ....132.1221H}.} \citep{2008A&A...489..651B}.  We also plot a 25,193 K blackbody \citep[the STIS determined temperature of Sirius B;][]{2005MNRAS.362.1134B} normalized to the V-band magnitude from STIS, the scaled mid-infrared excess of the prototype DAZd white dwarf, G29-38 (using the model SED from \citet{2005ApJ...635L.161R}, but ignoring the small-grain silicate dust feature), and the scaled mid-infrared excess of the hot white dwarf, SDSS 1228+1040 \citep[using the model SED of][]{2009ApJ...696.1402B}. Our mid-infrared data are consistent with the blackbody, and show no evidence of an infrared excess.

If Sirius B has no infrared excess (as appears to be the case, or is at least approximately true), our mid-infrared photometry fall on the Raleigh-Jeans tail for both Sirius A and B.  Thus, the $\Delta$ mag between Sirius A and B is expected to be the same in all of our mid-infrared filters.  As can be seen in Table \ref{Observations} and Figure \ref{Sirius B SED}, our data are not quite statistically consistent, which is likely due to residual correlated detector artifacts that have caused us to slightly underestimate our errors. This also explains why two of our filters (Si-2 and Si-4) appear to be statistically sub-photospheric in Figure \ref{Sirius B SED}.  To put a robust upper-limit on Sirius B's infrared excess, we take an unweighted average of our 5 mid-infrared photometry points, and find that the $\Delta$ mag between Sirius A and B is 10.77$\pm$0.09 mags.  This empirical error estimate implies that any infrared excess to Sirius B must be $\lesssim$10\% of the photosphere across the N-band.

The DAZd white dwarfs reported by \citet{2007ApJS..171..206M} and \citet{2007ApJ...662..544V}, and others, all have massive mid-infrared excesses \citep[on the order of $\sim$10,000\% at N-band for G29-38 as shown in Figure \ref{Sirius B SED} and][]{2005ApJ...635L.161R} due to a large quantity of small circumstellar dust grains.  Our non-detection of a mid-infrared excess around Sirius B clearly shows that it is not a member of this DAZd class of objects.\footnote{In comparing Sirius B to the DAZd class, we note that the DAZd white dwarfs described in the literature are mostly cooler (as described in Section 1) and do not have luminous binary companions.  In these white dwarfs, the infrared excess is probing dust very near the white dwarf (i.e. inside the tidal truncation radius).  In the case of Sirius B, which is hotter, dust within the tidal truncation radius is sublimated, but the outer regions of the disk are illuminated by Sirius A.}  However, the polluted atmospheres of DAZ white dwarfs only require a small amount (a Ceres-sized asteroid) of accreted mass to explain their spectra \citep{2010ApJ...722..725Z}, well below the detection/calibration limits of current telescopes.  Additionally, this mass can be hidden in cool material that is far enough from the white dwarf that it does not emit significantly in the mid-infrared.  A variety of novel observational techniques, such as interferometry, high-contrast imaging, and leveraging a nearby luminous companion to illuminate a large portion of the disk (as in the case of Sirius A) will be necessary to study the outer parts of these systems, which might comprise $\sim$1/4 of DA
white dwarfs, and $\sim$1/3 of DB white dwarfs \citep{1998ApJ...505L.143Z,2010ApJ...722..725Z}.

\section{Conclusions}

We used archival Gemini/T-ReCS data to directly image the nearest white dwarf, Sirius B, in 5 filters in the N-band (10$\micron$) window.  The data were taken over a several year period, where Sirius A was used as a photometric standard for many T-ReCS programs.  Because of Sirius B's non-negligable orbital motion during that timespan, we shifted each image by the binary's calculated orbital ephemeris in order to stack the images on Sirius B.

Although \citet{2008A&A...489..651B} reported a slight excess in the Ks-band, we find no evidence of a large mid-infrared excess, as would be expected for a DAZd white dwarf with a dusty debris disk \citep{2007ApJ...662..544V}.

White dwarfs in Sirius-like binary systems might be good targets for observing the outer parts of these debris disks.  Because of the low-luminosity of white dwarfs, circumstellar material at the separations typically observed in debris disks are too cold to emit significant radiation in the mid-infrared.  However, in binary systems, regions of the white dwarf's disk that are not heated by the white dwarf can be heated by its more luminous companion.

\acknowledgements
The authors thank the anonymous referee for his/her extremely helpful suggestions, and Eric Nielsen for providing us with his orbital ephemeris program.  AJS acknowledges the NASA Graduate Student Research Program (GSRP) for its generous support.  LMC is supported by an NSF CAREER award.

\clearpage

\begin{deluxetable}{ccccccccccccc}
\tabletypesize{\scriptsize}
\tablecaption{Gemini/T-ReCS Observations of Sirius A-B}
\tablewidth{0pt}
\tablehead{
\colhead{Filter} &
\colhead{Wavelength $\micron$\tablenotemark{a}} &
\colhead{On-Source Integration (s)} &
\colhead{$\Delta$ Mag (Sirius A-B)} &
\colhead{Sirius A Flux (Jy)\tablenotemark{b}} &
\colhead{Sirius B Flux (Jy)}
}

\startdata
Si-2 & 8.74  & 1639 & 10.91$\pm$0.09 & 168.75 & 0.0073$\pm$0.0006\\ 
Si-3 & 9.69  & 1014 & 10.58$\pm$0.23 & 140.18 & 0.0082$\pm$0.0018\\
Si-4 & 10.38 & 912  & 11.01$\pm$0.15 & 124.63 & 0.0049$\pm$0.0007\\
Si-5 & 11.66 & 2447 & 10.69$\pm$0.11 & 97.07 & 0.0051$\pm$0.0005\\
N   & 10.36 & 1752 & 10.67$\pm$0.15 & 131.12 & 0.0071$\pm$0.0010\\
\enddata
\tablecomments{Observations were taken sporadically from 2003-2006, as calibrations for other Gemini programs, for which Sirius A is a common photometric/PSF standard.}
\tablenotetext{a}{Filter information available at http://www.gemini.edu/sciops/instruments/T-ReCS/imaging/filters.  The Si filters are $\sim$10\% bandwidth while the N filter is $\sim$50\% bandwidth.} 
\tablenotetext{b}{Sirius A fluxes through standard filters available at
http://www.gemini.edu/sciops/instruments/mir/MIRStdFluxes.txt}

\label{Observations}
\end{deluxetable}

\clearpage

\clearpage

\clearpage

\clearpage

\bibliographystyle{apj}
\bibliography{database}

\end{document}